\pdfoutput=1
\documentclass[aps,prb,twocolumn]{revtex4}
\usepackage{amssymb}
\usepackage{graphicx}
\usepackage{amsmath}
\usepackage{amsfonts}
\usepackage{amssymb}%

\begin{document}
\title{Chemically-exfoliated single-layer MoS$_2$ : stability,\\ 
lattice dynamics and catalytic adsorption from first principles }

\author{Matteo Calandra}
\email[]{matteo.calandra@upmc.fr}
\affiliation{Universit\'{e} Pierre et Marie Curie, IMPMC, CNRS UMR7590, 4 Place Jussieu, 75005 Paris, France}

\begin{abstract}
Chemically and mechanically exfoliated MoS$_2$ single-layer samples have
substantially different properties. While mechanically exfoliated
single-layers are mono-phase ( 1H polytype with Mo in trigonal prismatic
coordination), 
the chemically exfoliated
samples show coexistence of three different phases, 1H, 1T (Mo in octahedral coordination) and
1T$^{'}$ (a distorted  $2\times 1$ 1T-superstructure). By using first-principles 
calculations, we investigate the energetics and the dynamical stability of the three phases. 
We show that the 1H phase is the most stable one, while the 
metallic 1T phase, strongly unstable, undergoes a phase
transition towards a metastable and insulating 1T$^{'}$ structure composed of separated
zig-zag chains. We calculate electronic structure, phonon dispersion, Raman frequencies
and intensities for the 1T$^{'}$ structure. We provide a microscopical description
of the J$_1$, J$_2$ and J$_3$ Raman features first detected more then $20$ years ago, but
unexplained up to now.
Finally, we show that H adsorbates, that are naturally present at the
end of the chemical exfoliation process, stabilize the 1T$^{\prime}$ 
over the 1H one. 
\end{abstract}

\maketitle

\section{Introduction}
Bulk transition metal dichalcogenides (TMD) are layered van der Waals
solids displaying remarkable properties promising both for fondamental research
as well as for technological applications. 
Metallic bulk transition
metal dichalchogenides (TMD) like NbSe$_2$, 
present coexistence of charge density wave and 
superconductivity \cite{DiSalvo,IwasaMoS2}, while insulating TMD (MoS$_2$,
WS$_2$) are flexible, have high mobilities and are routinely
used in flexible electronics.

Since the pioneering work of Frindt and coworkers \cite{ FrindtJAP1966,
  JoensenMRB1986, FrindtPRL1972, YangPRB1991} and the successive
developments in the fields of mechanical \cite{NovoselovPNAS} and
liquid \cite{ColemanSCI} exfoliation, it has been possible to obtain
free-standing or supported single-layer TMD. These monolayers
are the inorganic analogue of graphene and display a rich 
chemistry\cite{ChhowallaNChem} that makes them attractive 
for energy storage applications.
Insulating single-layer TMD have much lower
mobilities \cite{Radisavljevic2011, FuhrerComment} than 
Graphene, but are nevertheless interesting for 
nanoelectronics, mainly due to the presence of a finite
bandgap.  

In this context, MoS$_2$ is considered one of the most
promising materials\cite{NotAlone}. The most stable polytype of bulk MoS$_2$
is the 2H (Molybdenite), where each Mo has a trigonal prismatic coordination
with the nearby S atoms. Mechanical exfoliation of bulk 2H MoS$_2$ 
lead to formation of single layer samples with the same local
coordination (here labeled 1HMoS$_2$). 
In chemically exfoliated samples the situation is different. 
In the first step of chemical exfoliation of bulk 2H MoS$_2$, 
Li atoms are intercalated between the layers. 
The Li intercalation stabilize a 1T Li$_{x}$MoS$_2$
polytype having each Mo octahedrally coordinated with the nearby S
atoms. Subsequent hydratation with excess water and ultrasonication leads to the separation
of the layers via LiOH formation and synthesis of large-area single-layer MoS$_2$ samples
\cite{JoensenMRB1986}. 
\begin{figure*}
\begin{minipage}[c]{0.2\linewidth}
\includegraphics[scale=0.125,angle=0]{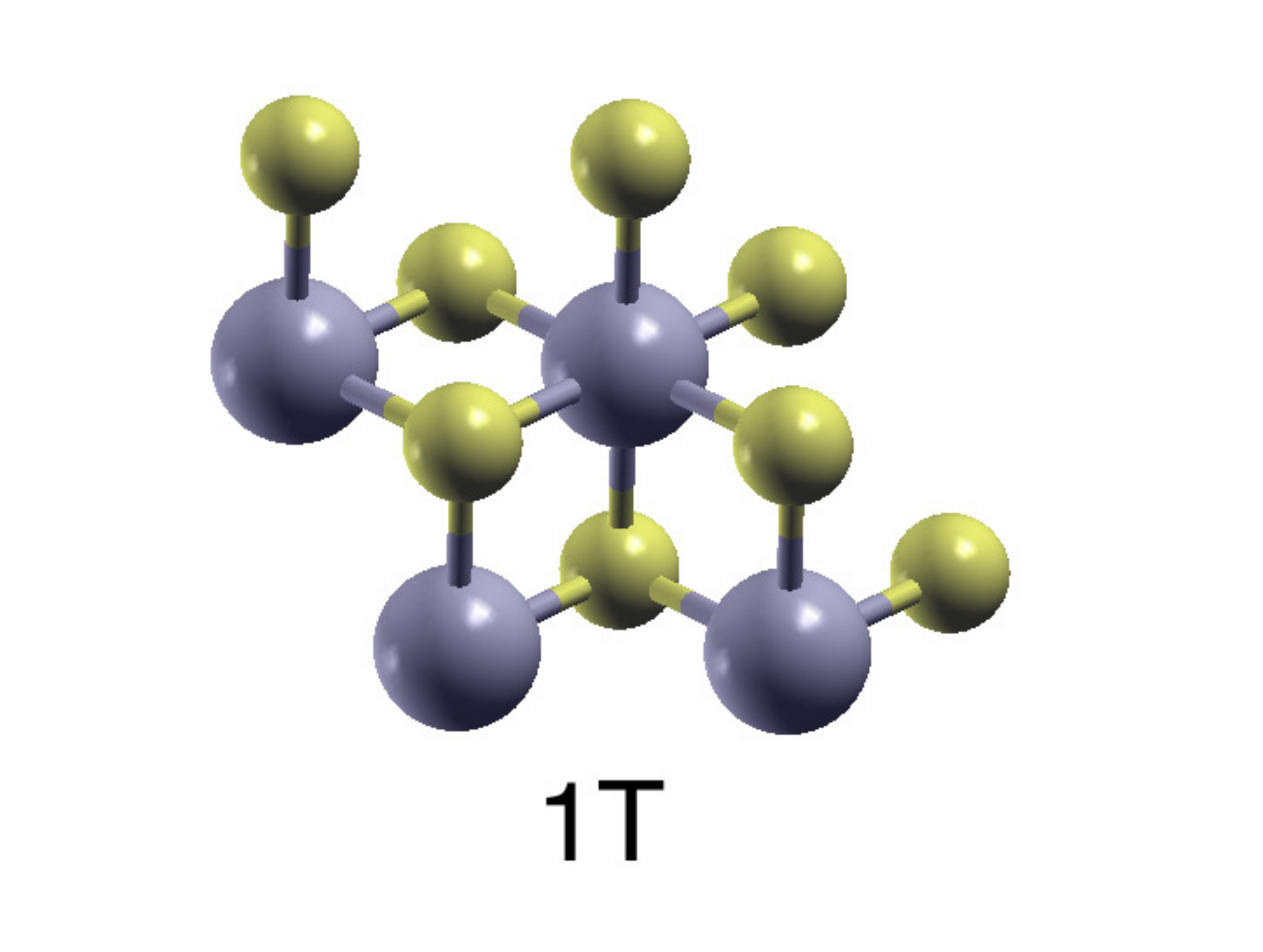}
\includegraphics[scale=0.125,angle=0]{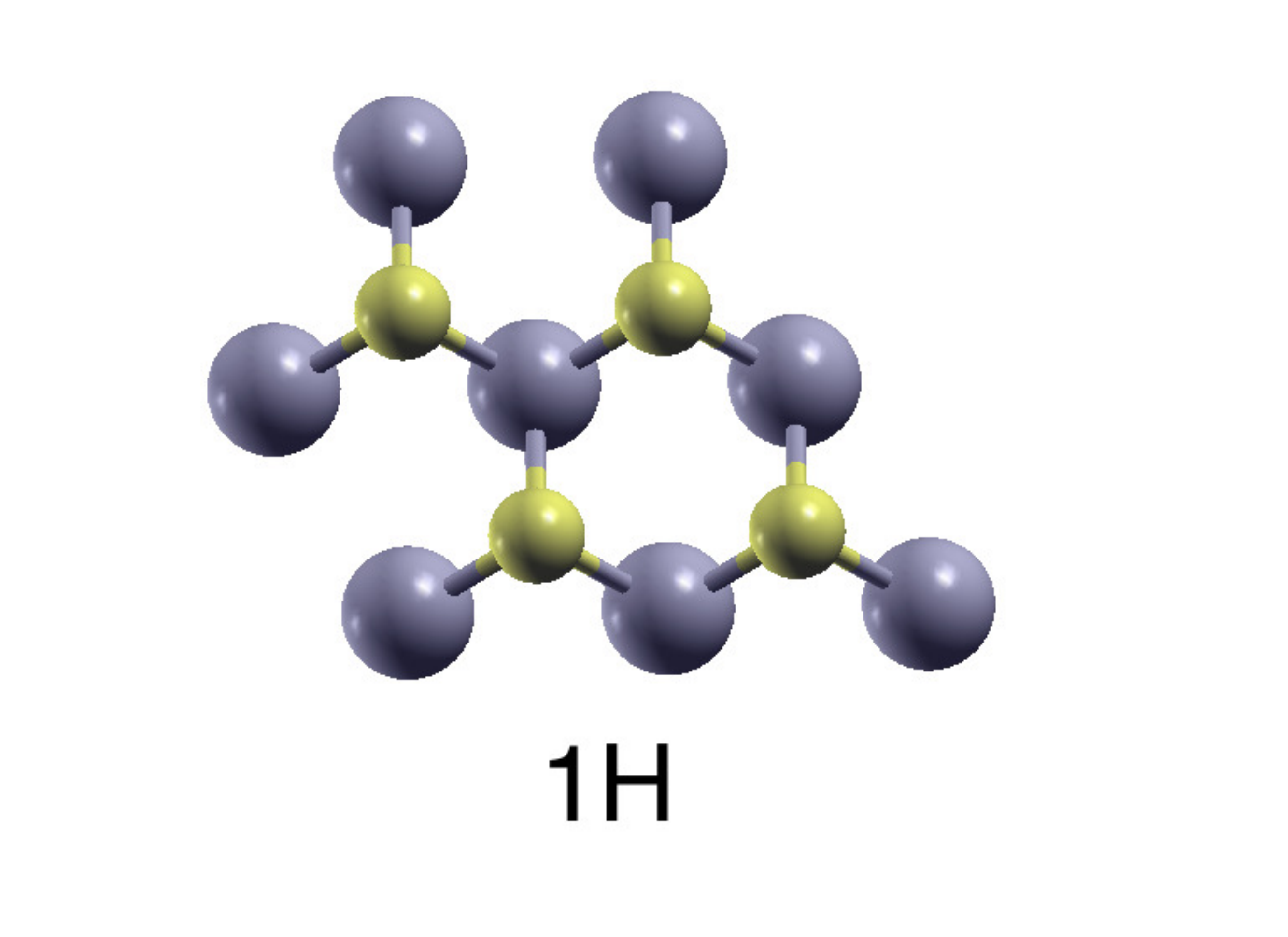}
\end{minipage}
\begin{minipage}[c]{0.7\linewidth}
\includegraphics[scale=0.225,angle=0]{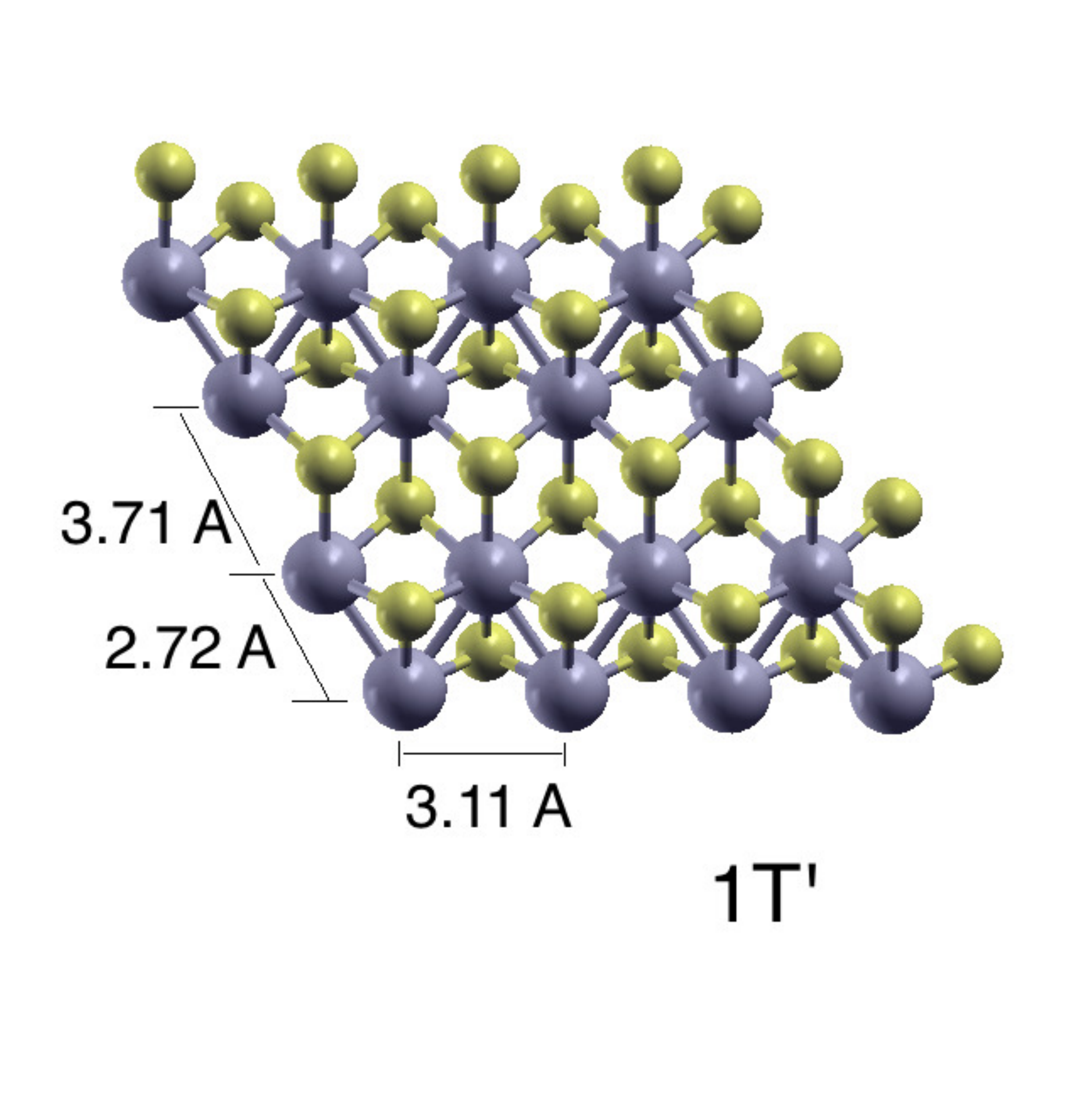}\includegraphics[scale=0.12,angle=0]{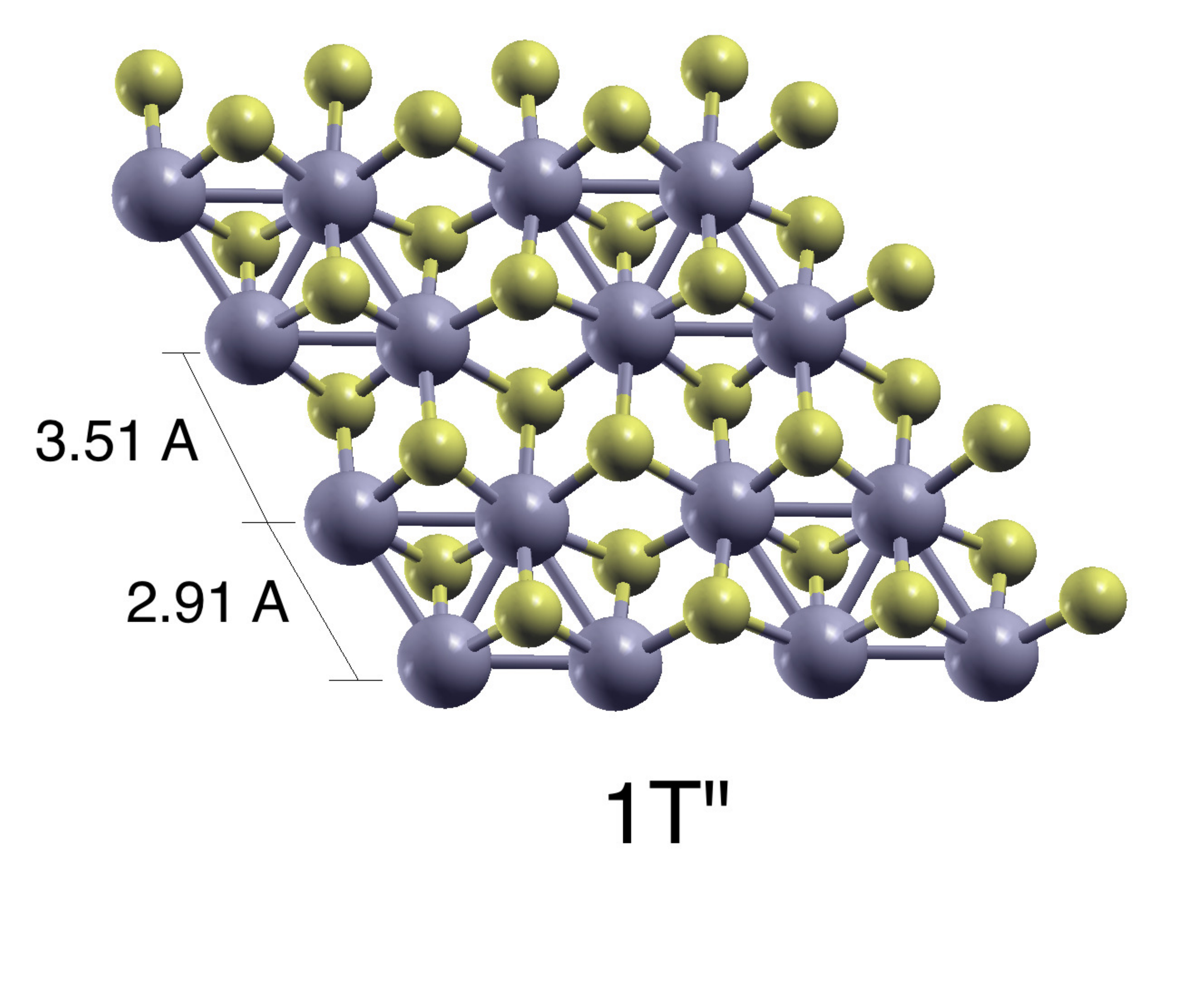}
\end{minipage}\hfill
\caption{Phases of chemically exfoliated MoS$_2$. The
1H has trigonal prismatic coordination and is the most stable
among all polytypes.
The 1T, 1T$^{'}$ and 1T$^{''}$ polytypes all have octahedral
coordination.  The   1T$^{'}$ is the lowest energy polytype among
those with octahedral coordination. The in-plane Mo-Mo distance is
$3.193$ \AA\, and $3.183$ \AA\, for the 1T$^{'}$ and 1H structures, respectively}
\label{fig:struct}
\end{figure*}

The properties of chemically exfoliated MoS$_2$ single layers are poorly
understood. Recently it has been shown that 
these samples are actually composed of heterostructures 
of 1H, 1T and 1T-distorted (labeled 1T$^{\prime}$)
MoS$_{2}$ phases \cite{EdaACSnano}. The 1T$^{\prime}$ phase is a $2\times 1$
superstructure of the 1T phase formed by zig-zag chains. Remarkably, the
three phases cohexist in the same sample and have substantially different
conducting properties as the 1T phase is metallic while the 1H and
1T$^{\prime}$ are insulating\cite{EdaNanoL}. Upon mild annealing at $200-300$ C
the 1T and 1T$^{\prime}$ phase disappear and transform
in the 1H one. 
Exposure to a $60-80$ keV electron beam induces S vacancies
\cite{Komsa} and transforms the 1T$^{\prime}$ phase into
the 1T one\cite{EdaACSnano,Suenaga}. Finally, it is important to remark that
chemically exfoliated single layers are covered
with absorbates that can play some role in stabilizing one structure or
the other.

The dynamical properties of the 1T$^{\prime}$ phase are not
understood. 
For example, while it is well established that the high energy optical Raman
spectra of the 1H phase are composed of two prominent peaks,
attributed to the E$_{\rm 2g}$ mode at $\approx 385 $ cm$^{-1}$ and to
the A$_{\rm 1g}$ mode at $\approx 403$ cm$^{-1}$\cite{Lee}, little is known on
the Raman spectra  of the 1T$^{\prime}$ phase. Raman
measurements \cite{Sandoval} on freshly-prepared single-layers with dominant 1T$^{\prime}$ phase 
show that the E$_{\rm 2g}$ peak  is missing, while at least five additional peaks
appear at lower energies (some of
these peaks are labeled J$_1$, J$_2$, J$_3$). Nothing is known on the phonon displacements
generating these features.

In this work we study the stability, the electronic structure and the
dynamical properties of the 1T and 1T$^{\prime}$ phases in single
layer MoS$_2$ by using density functional theory (DFT) calculations. 
We show that the metallic 1T phase is dynamically
unstable. We find distorted structures with a $2\times 1$
(1T$^{'}$MoS$_2$) and $2\times 2$ (labeled
1T$^{''}$MoS$_2$) real space
periodicities having lower energies then the 1T one. 
Both  1T$^{'}$ and 1T$^{''}$ structures are, however,
substantially higher in energy then the 1HMoS$_2$ phase (see Fig.\ref{fig:struct}  for 
a plot of the crystal structure of the different phases).

We then fully characterize the 
distorted 1T$^{\prime}$ phase found in experiments on chemically
exfoliated MoS$_2$, by obtaining
its electronic structure, phonon dispersion and Raman intensities.
Finally, we study catalytic absorption in the 1T$^{'}$ phase and show
that H adsorbates stabilize the 
1T$^{\prime}$ with respect to all the others.

The paper is organised as follows. In section \ref{sec:tech} we
describe the technical details of the calculation. In
sec. \ref{sec:oct}
we analyze the stability of octahedral phases with respect to the
trigonal prismatic ones and in sec. \ref{sec:raman} we study the Raman
spectrum of the distorted 1T$^{'}$ phase. Finally, in sec.
\ref{sec:cata} we study catalytic adsorption of Hydrogen and its
effect on the structural and electronic properties of the different
structures. 

\section{Technical details\label{sec:tech}}
The results reported in the present paper were obtained from
first-principles 
density functional theory in the generalized gradient approximation\cite{PBE}.
The  QUANTUM-ESPRESSO\cite{QE} package was used with norm-conserving
pseudopotentials and a plane-waves cutoff energy of $90$ Ry. Semicore
states were included in the Mo pseudopotential. The 
Electronic structure calculations were performed 
by using a $24\times
24 $, $12\times 24$, $12\times 12$ electron-momentum grids for the
1T, 1T$^{'} $ and 1T$^{''}$ phases, respectively. For the metallic 1T structure
we use an Hermitian-Gaussian smearing of $0.01$ Ryd.
The phonon dispersion of the 1T phase was calculated by Fourier
interpulating dynamical matrices calculated on a $8\times 8$
phonon-momentum grid and on a $24\times
24$ electron-momentum grid. The Raman intensity calculation for the
1T$^{\prime}$ phase was performed on a $8\times 16$ electron-momentum
grid.
The phonon dispersion calculation for the 1T$^{'}$ structure was
performed using a $4\times 4$ phonon momentum grid. 

\section{Results}

\subsection{Stability of octahedral phases\label{sec:oct}}

\begin{figure}
\centerline{\includegraphics[scale=0.5,angle=0]{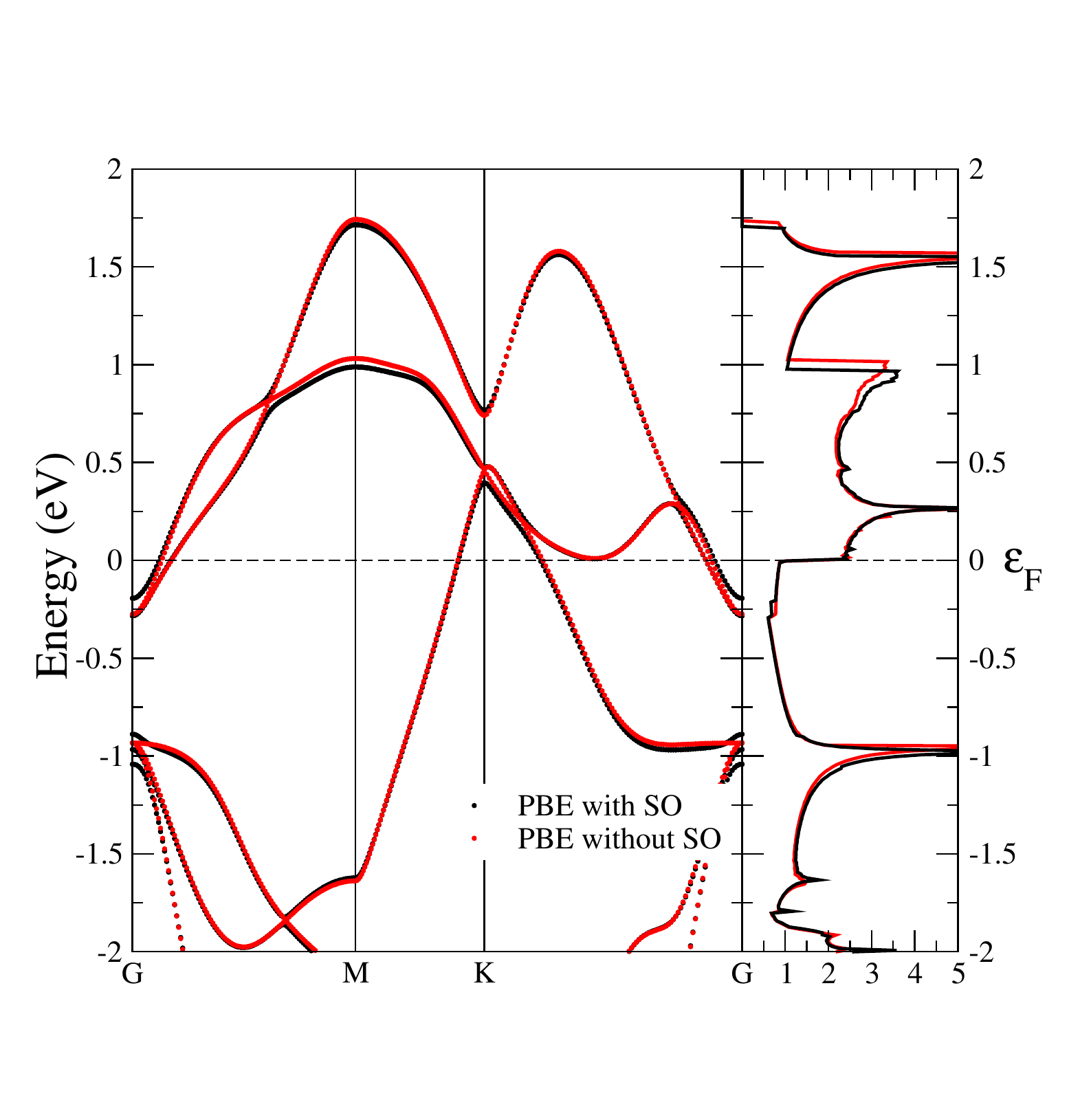}}
\caption{Electronic structure and density of states of the
1TMoS$_2$ phase with or without spin orbit coupling. The energy 
are plotted with respect to the Fermi level.  } 
\label{fig:1Tpure_el}
\end{figure}

We first investigate the relative stability of 2H and 1T phases in Fig. \ref{fig:struct}.
As expected, we find that the 1H MoS$_2$ phase is the most stable one,
with a lower energy of 0.83 eV / Mo atom with respect to the 1T one.
The electronic structure calculation of the 1T
structure  in Fig. \ref{fig:1Tpure_el} shows that this polytype is indeed metallic. 
Differently from the 1H case, here the spin-orbit coupling
is very weak and from now on it will be neglected. 
 
\begin{figure}
\centerline{\includegraphics[scale=0.5,angle=0]{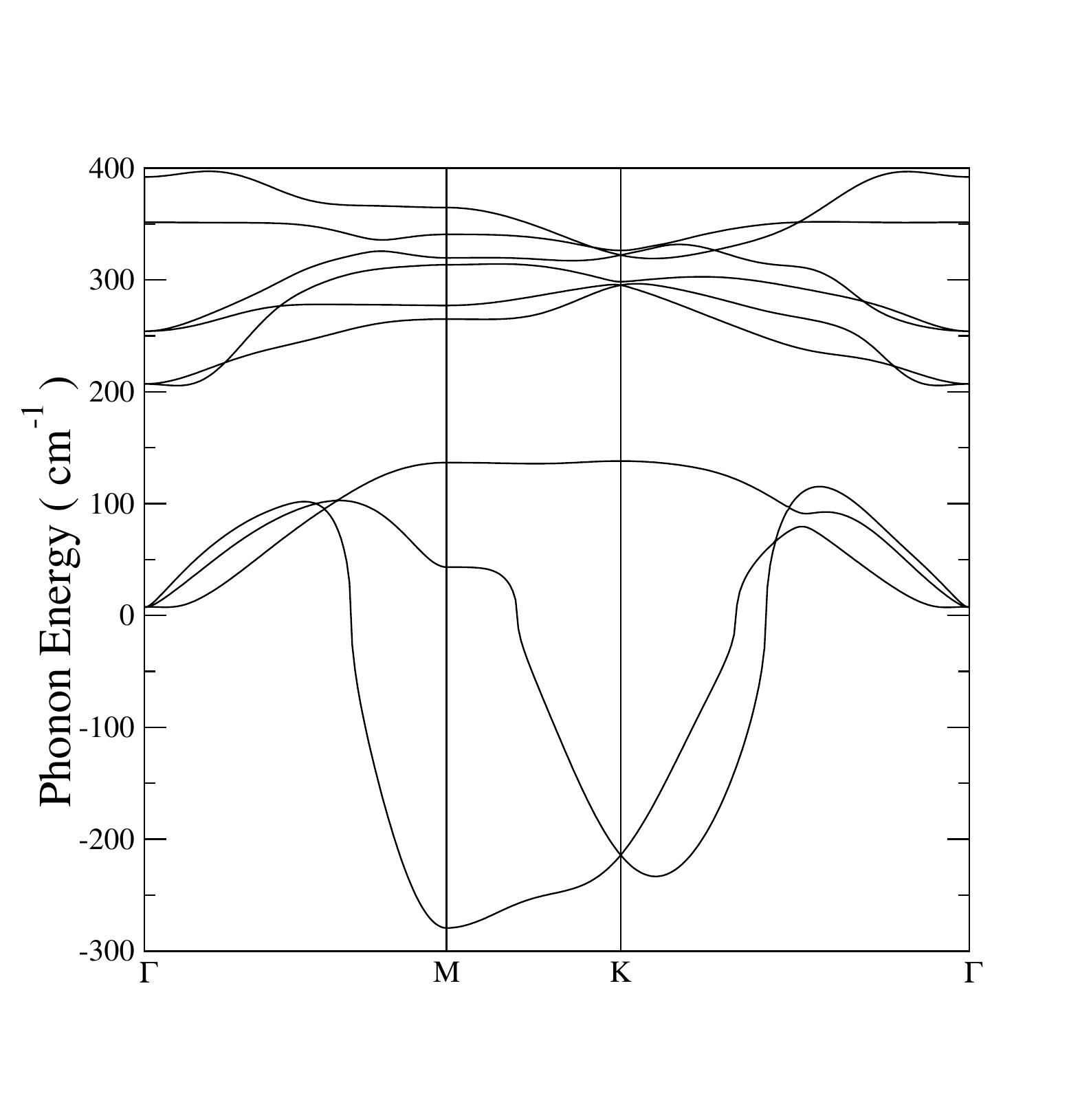}}
\caption{Phonon dispersion of the 1TMoS$_2$ phase 
showing a dynamical instability at zone border.} 
\label{fig:1Tpure_ph}
\end{figure}

\begin{figure}
\includegraphics[scale=0.5,angle=0]{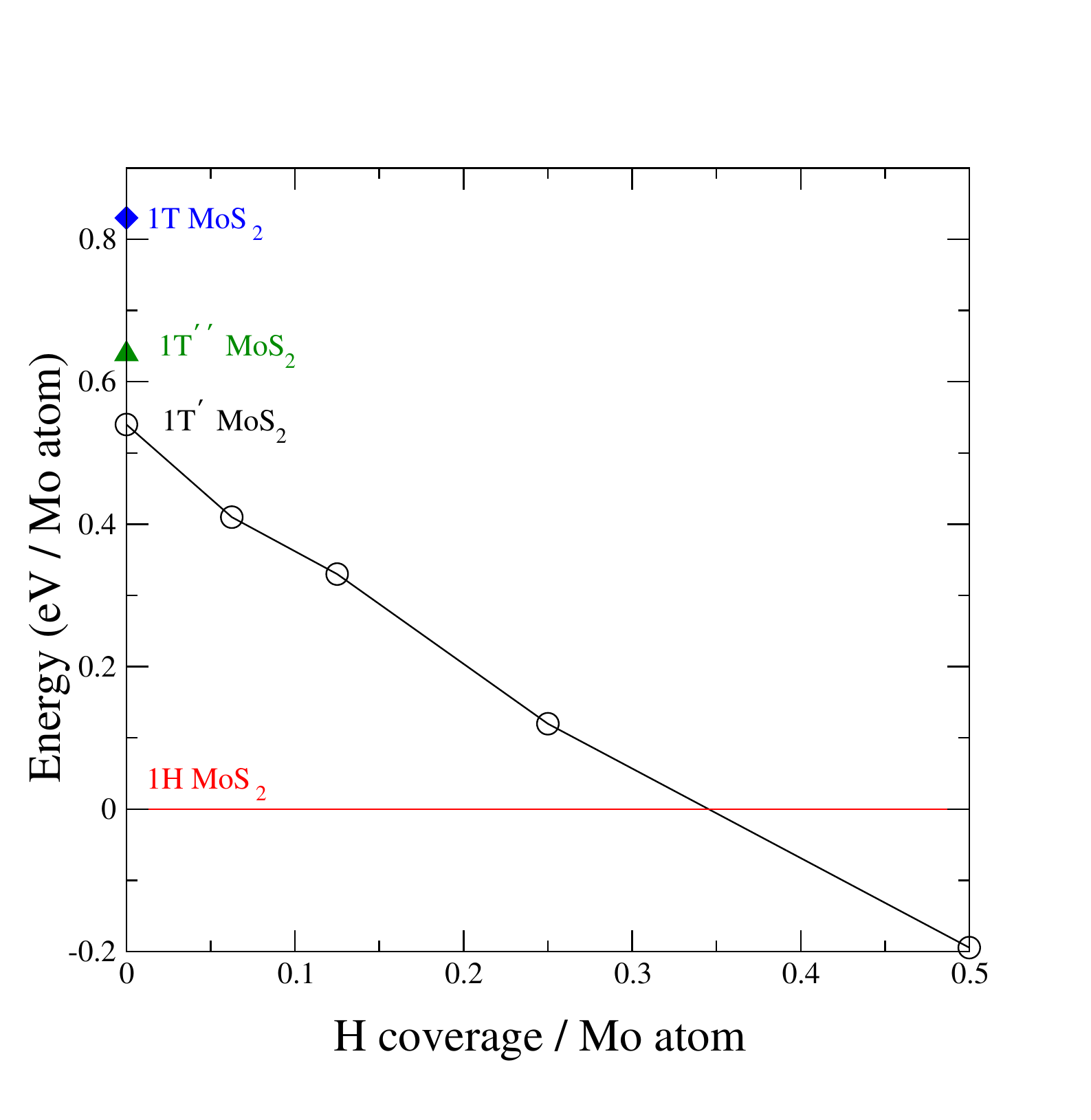}
\caption{Stability of different MoS$_2$ structures with respect to the
1H polytype and as a function of H coverage per Mo atom}
\label{fig:H_Stability}
\end{figure}

As the energy difference between the 1T and 2H phases is 
more then 30 times larger then the 200-300 K  
annealing temperature necessary to transform the 1T phase
in the 2H one, the experimental detection of the 1T and 
1T$^{\prime}$ phases cannot be inferred from the total energy
difference between the two.
It has been suggested that the 1T phase is metastable and, as a consequence, an
energetic barrier occurs between the two \cite{ChhowallaNChem}.
To verify this hipothesis, we calculate the phonon dispersion
for the 1T phase. We find that the 1T structure is dynamically
unstable (see Fig. \ref{fig:1Tpure_ph}) at zone border, with the largest
instability at the M point of the hexagonal lattice. 
\begin{table}[hb!]
\begin{tabular}{l c c c }
\hline
Atom &    x      &       y    &      z      \\
\hline
Mo  &  0.0508948 &  0.0508948 &  0.0051972  \\ 
S   &  0.1662841 &  0.6662835 &  0.1240922  \\
S   &  0.3337158 &  0.3337161 & -0.1240922  \\
Mo  &  0.4491051 & -0.0508948 & -0.0051972  \\
S   &  0.6714116 &  0.6714108 &  0.0957802  \\
S   &  0.8285883 &  0.3285888 & -0.0957802  \\
\hline
\end{tabular}
\caption{Atomic coordinates with respect to the direct axis for the 1T$^{'}$ structure. The lengths of
 the two direct lattice vectors of the two dimensional lattice are identified by
$a=6.411 \AA$, $b=3.111 \AA$. The angle between them
$\gamma=119.034^{\circ}$. \label{tab:structTprime}}
\end{table}
This distortion is compatible with a $2\times 1$ superstructure.
To identify the lowest energy superstructure, we perform calculations on a $2\times 1$
supercell, by displacing the atoms along the direction given by the
phonon displacement of the most unstable mode at M. We find that substantial energy 
( 0.19 eV/ Mo) is gained by the distortion. We then start from this
distorted structure and perform full structural optimization 
of internal coordinates and of the 2D-cell. As shown in
Figs. \ref{fig:H_Stability},  we find a stabilization
of  an octahedrally coordinated structure composed of zig-zag chains, with an energy
gain of 0.29 eV/Mo with respect to the 1T phase. 
Structural parameters of the zig-zag distorted-structure are given in
Tab. \ref{tab:structTprime}.
Here we remark that the
shortest distance between Mo atoms belonging to the same chain is $\approx
2.72 $\AA , while the shortest distance between atoms on different
chains is $\approx 3.71 $\AA . 
The angle between the Mo atoms
in the chain is $\approx 69.64^{\rm o}$. 
The in-plane nearest-neighbours Mo-Mo distance of the 1T$^{'}$ structure is 
almost identical to the nearest neighbours
distance of Mo atoms in bcc Mo\cite{Ashcroft}, that is $2.728$\AA. 
On the contrary in the 1T structure the Mo-Mo bonds is $3.193$\AA,
 substantially elongated with respect to the Mo-Mo nearest neighbout
distance in bcc Mo.

The devised 1T$^{'}$ structure closely
resembles that detected in experiments on chemically exfoliated
samples \cite{EdaACSnano}. 

\begin{figure}
\includegraphics[scale=0.5,angle=0]{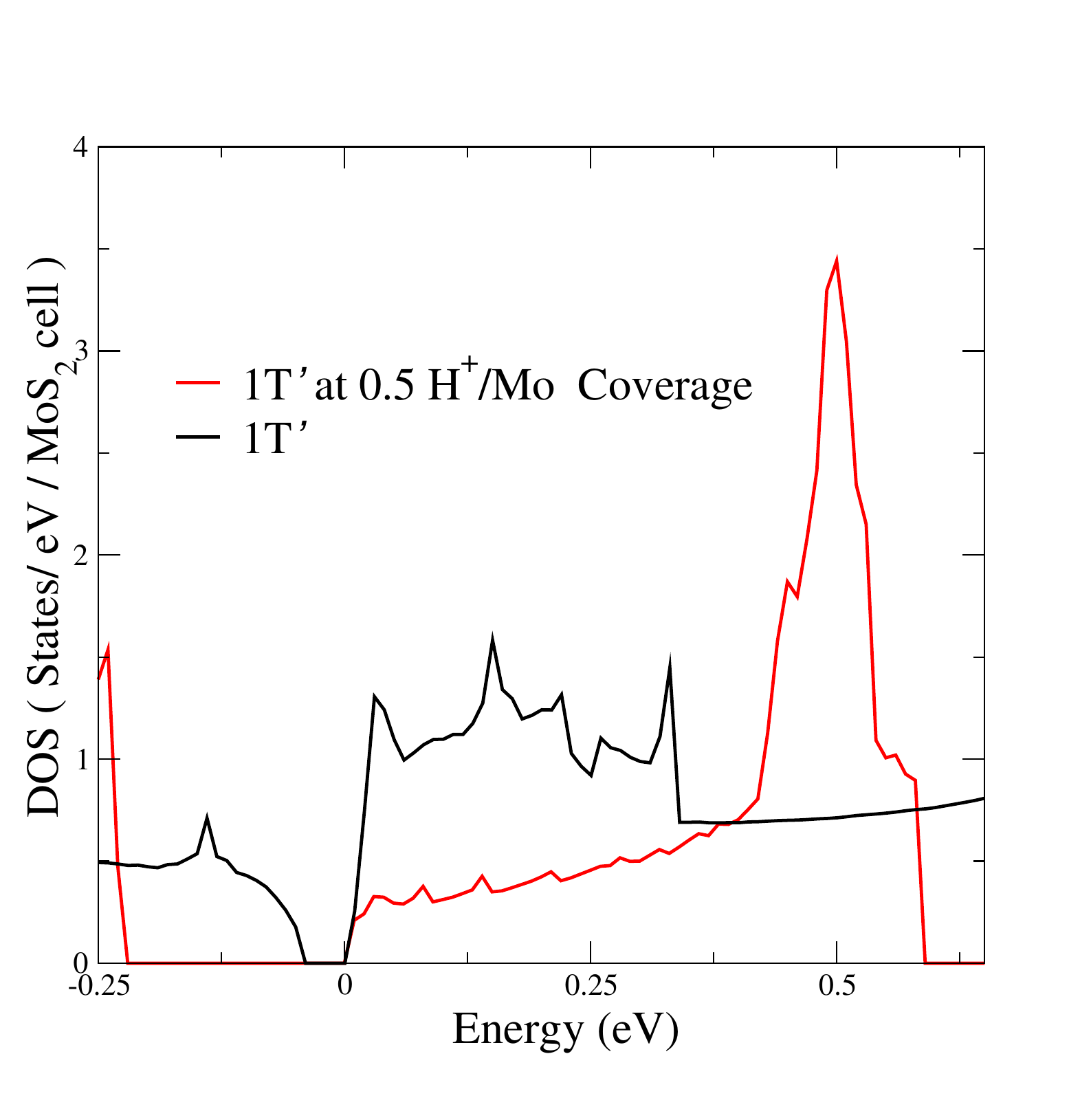}
\caption{ Electronic density of states of the 1T$^{\prime}$ at 0 and 0.5 H/Mo
coverage. The zero of the energy has been set to the bottom of
the conduction band.}
\label{fig:Dos_Distorted}
\end{figure}

The electronic density of states of the distorted structure is shown
in Fig. \ref{fig:Dos_Distorted}. The distortion opens a very small gap
($\approx 0.045$ eV) that makes the system insulating. The formation
of zig-zag chains is actually very similar to the standard
Peierls dimerization in one dimensional systems, i. e.  the system
gains energy in opening a gap. The Peierls distortion reduces the
dimensionality of the 2D layer that is now broken in 1D zig-zag chains.
This is at odd with most bulk metallic 
transition metal dichalcogenides where the charge density wave state
coexists with metallicity and superconductivity \cite{DiSalvo}.
However, given the large energy gain and the strong bond
deformation involved in this distortion, the transition to 1D-zig-zag
chains has to be considered more a real structural transition then a
charge density wave.

\begin{table}[hb!]
\begin{tabular}{l c c c }
\hline
Atom &    x      &       y    &      z      \\
\hline
Mo   &      0.022531337  &   0.022531337  &   0.0    \\
S    &      0.317728127  &   0.651403657  &   0.056424736    \\
S    &      0.651403657  &   0.317728127  &  -0.056424736    \\
Mo   &      0.465000117  &   0.001783220  &  -0.000455305    \\
S    &      0.815301793  &   0.651805630  &   0.049761398    \\
S    &      1.150643861  &   0.316295157  &  -0.062393003    \\
Mo    &     0.444399776  &   0.444399776  &   0.0   \\
S     &     0.814643878  &   1.148463448  &   0.056499460   \\
S     &     1.148463448  &   0.814643878  &  -0.056499460   \\
Mo    &     0.001783220  &   0.465000117  &   0.000455305   \\
S     &     0.316295157  &   1.150643861  &   0.062393003   \\
S     &     0.651805630  &   0.815301793  &  -0.049761398   \\
\hline
\end{tabular}
\caption{Coordinates of the 12 atoms in the 1T$^{''}$ unit-cell with
respect to the direct lattice vectors The lengths of the two direct
lattice vectors of the two dimensional lattice are identified by
$a=b=6.422 \AA$. The angle between the two is $\gamma=119.331^{\circ}$.\label{tab:structTdec}}
\end{table}
As the energy difference between the 1T$^{'}$ and the 1H structures is
large ( 0.54 eV/Mo ), we perform additional structural optimization on the
$2\times 2$ supercell to see if other superstructures can be
stabilized. We do indeed find another distorted structure formed
by Mo rhombus (1T$^{''}$ MoS$_2$, see Fig. \ref{fig:struct} and Tab.\ref{tab:structTdec}) 
that is 0.19 eV lower in energy then the 1T structure
but still higher then both the 1T$^{'}$ and the 1H one.
Interestingly, in past experimental works on chemically exfoliated MoS$_2$ samples
\cite{YangPRB1991}, a similar 1T$^{''}$ structure was proposed as the most stable one in
the monolayer.

\subsection{Raman spectra of the distorted 1T' phase \label{sec:raman}}

In order to
substantiate that the 1T$^{\prime}$ structure determined theoretically
is the same of the experimental one, we calculate the phonon
frequencies at zone center and the first-order Raman intensities for the 1T$^{\prime}$
structure. We also give a complete interpretation of Raman spectra
in chemically exfoliated samples that is currently lacking in
literature.
\begin{table}
\begin{tabular}{cccc} 
{\it Theory }    & {\it Theory} &  {\it   Experiment}\cite{Sandoval}   & 1HMoS$_2$  \cite{Lee}     \\ 
{\it (cm$^{-1}$)} & {\it (Intensity)}  & {\it (cm$^{-1}$)} & {\it (cm$^{-1}$)}  \\ \hline
  147       &       0.003     &                    &
\nonumber \\
   151       &       0.008     &  156   (J1)    & 
\nonumber \\
   216       &       1.0         &  226   (J2)    &
\nonumber \\
   223       &       0.006     &                    &
\nonumber \\
   286       &       0.011     &  287            &
\nonumber \\
   333       &       0.033     &  333   (J3)    &
\nonumber \\
   350       &     $<$0.001    & 358               &   385 (E$_{2g}$)    
\nonumber \\
   412       &       0.13       &  408            &   403 (A$_{1g}$)
   \\ \hline
\end{tabular}
\caption{Calculated phonon frequencies  (in
cm$^{-1}$) and first-order Raman intensities of the 1T$^{\prime}$ phase, as compared
with experiments on both 1T$^{\prime}$ and 1H phases. The intensities
are normalized to the most intense peak. The incoming and outcoming
light in the Raman experiment are assumed to be unpolarized. See Ref. \onlinecite{Boukhicha}
for more details on the definition of the Raman intensities.\label{tab:Raman}}
\end{table}

\begin{figure*}
\includegraphics[scale=0.125,angle=0]{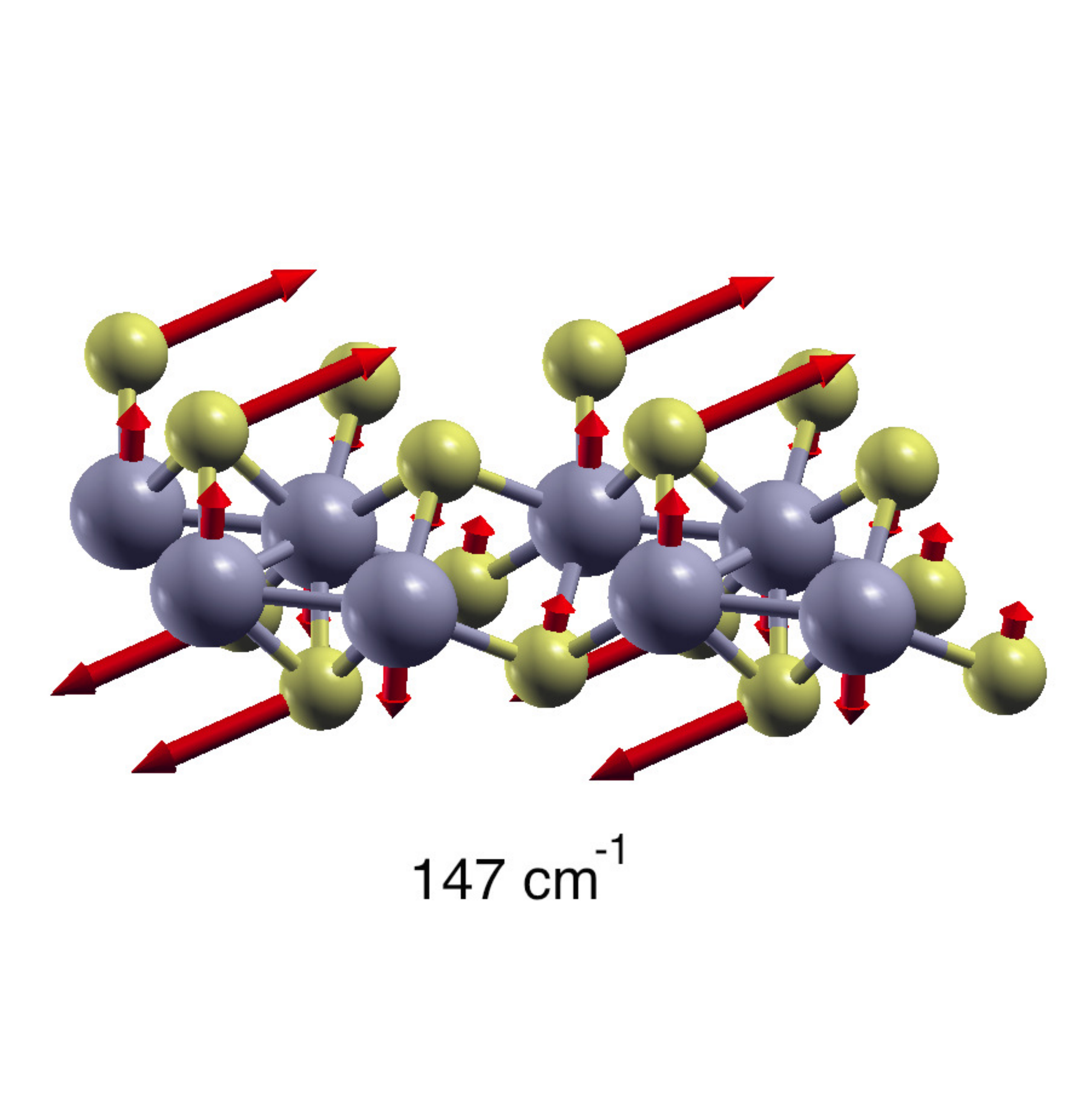}\includegraphics[scale=0.125,angle=0]{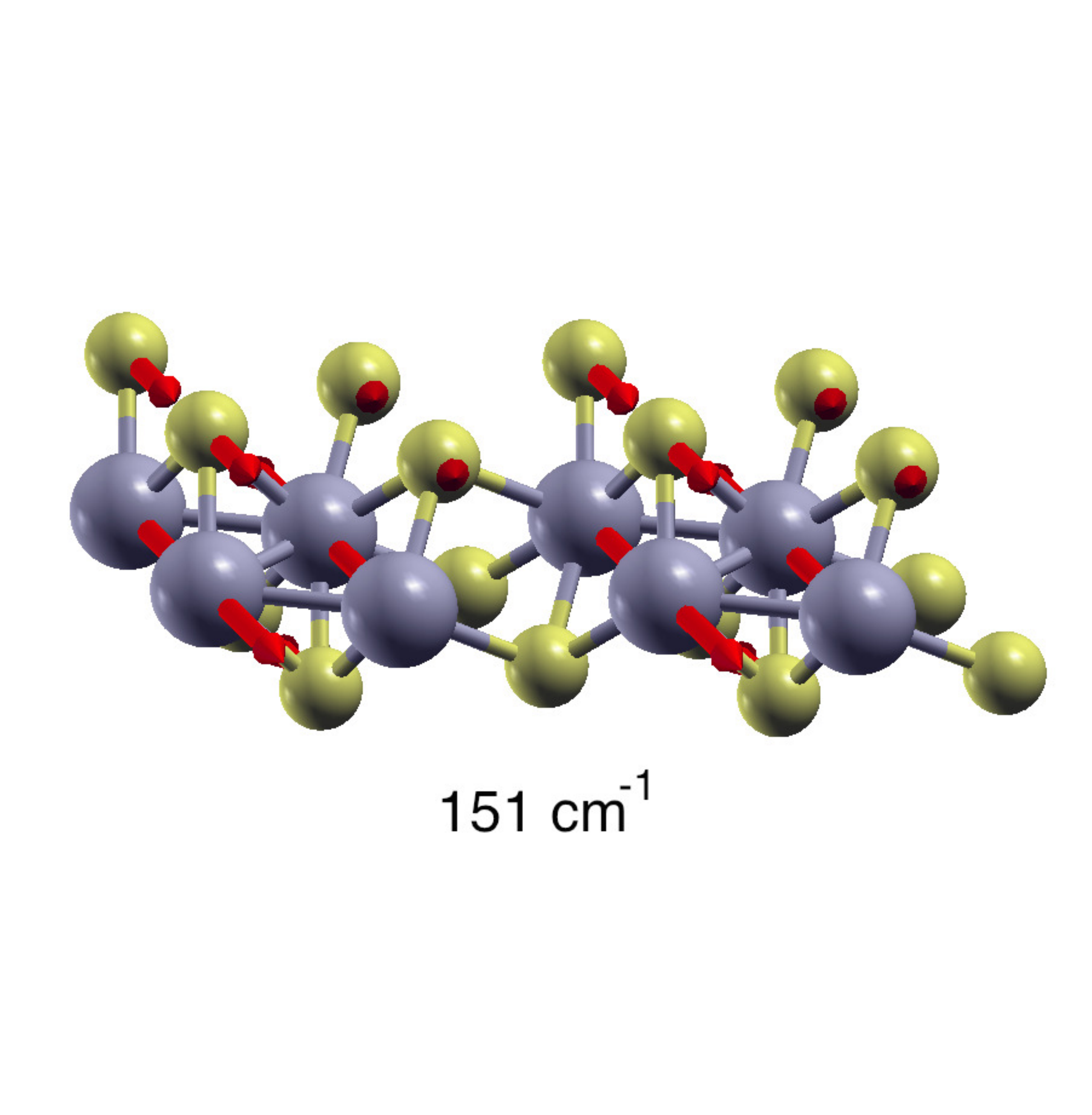}
\includegraphics[scale=0.125,angle=0]{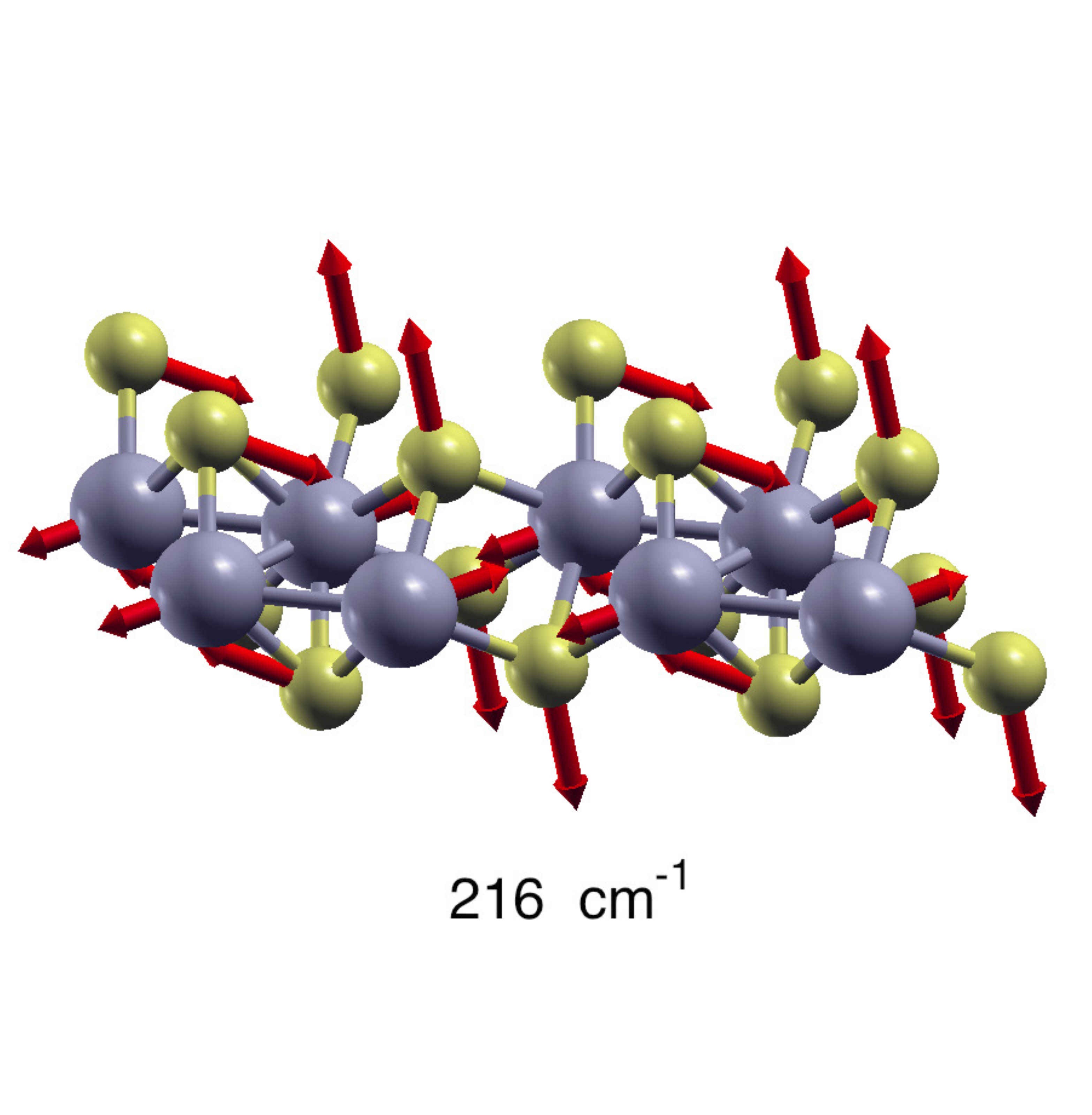}\includegraphics[scale=0.125,angle=0]{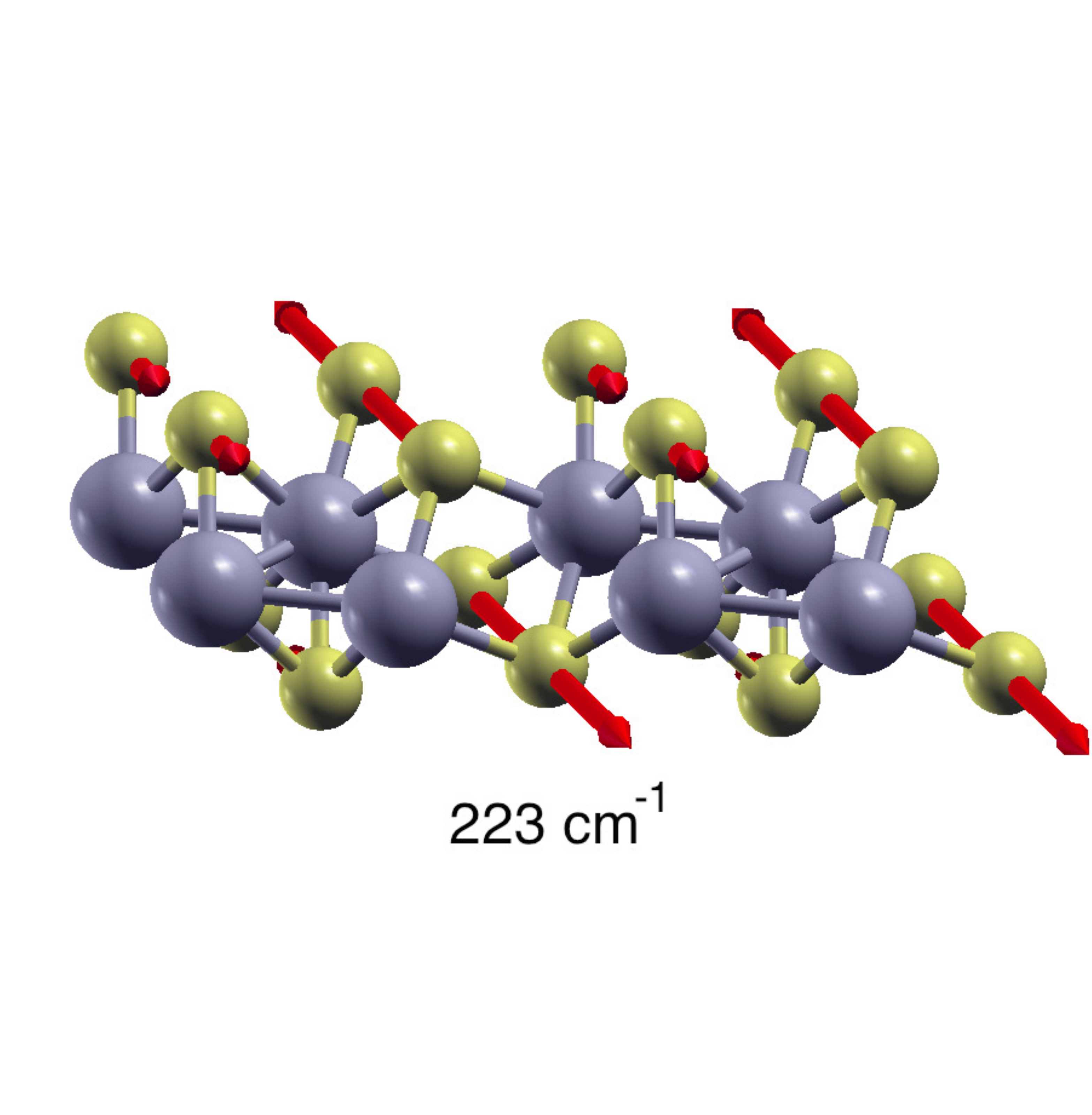}
\includegraphics[scale=0.125,angle=0]{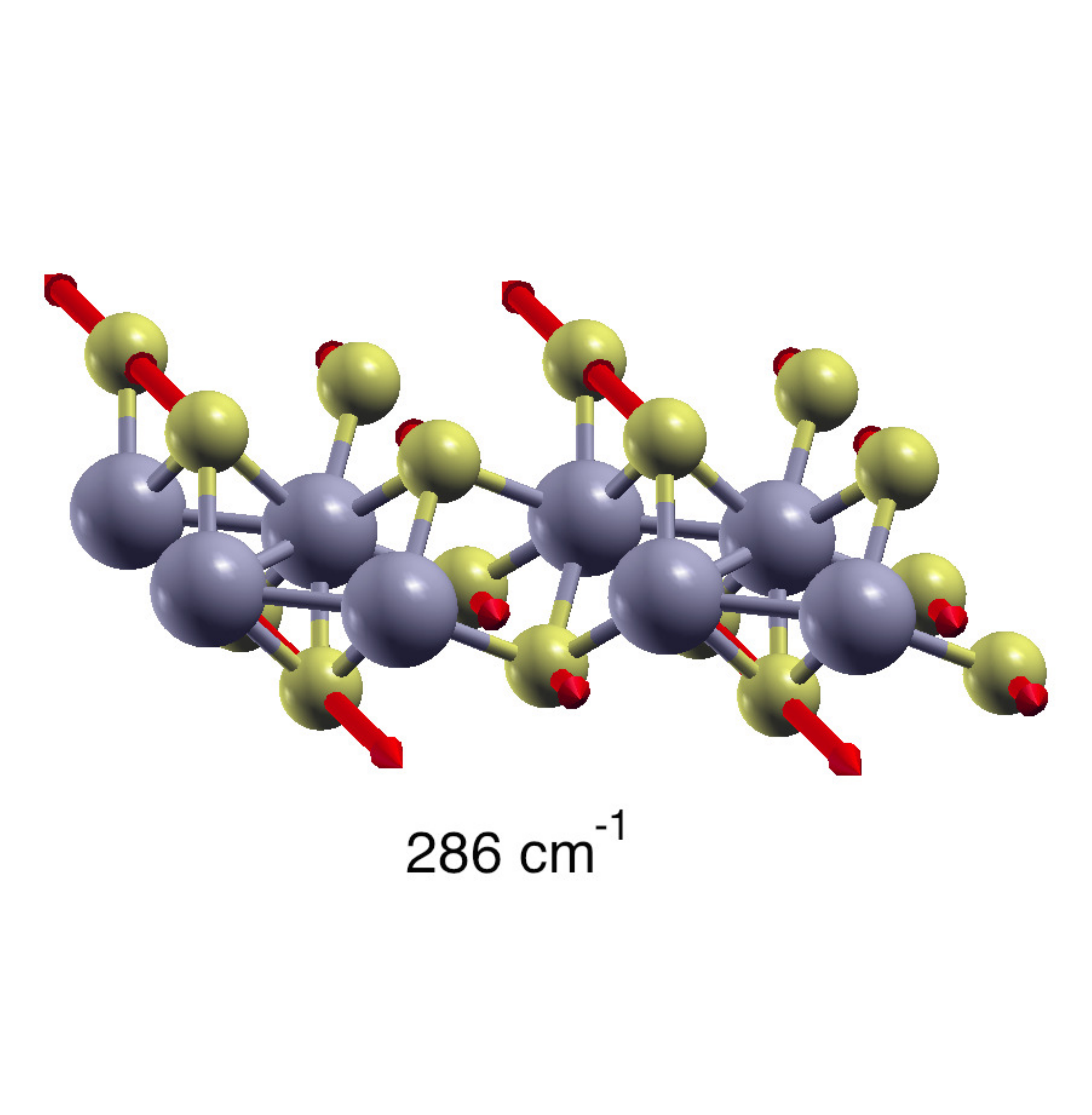}\includegraphics[scale=0.125,angle=0]{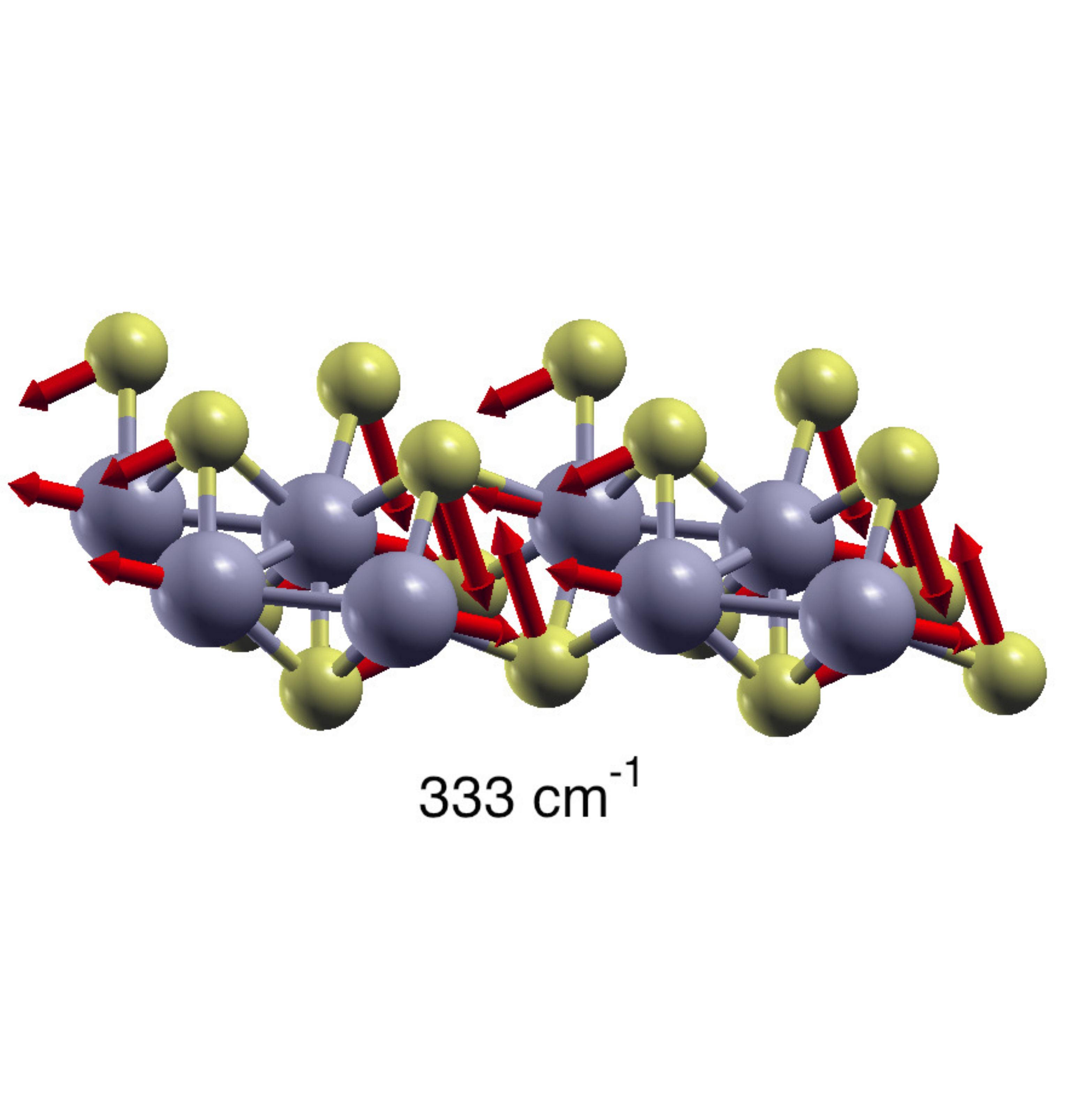}
\includegraphics[scale=0.125,angle=0]{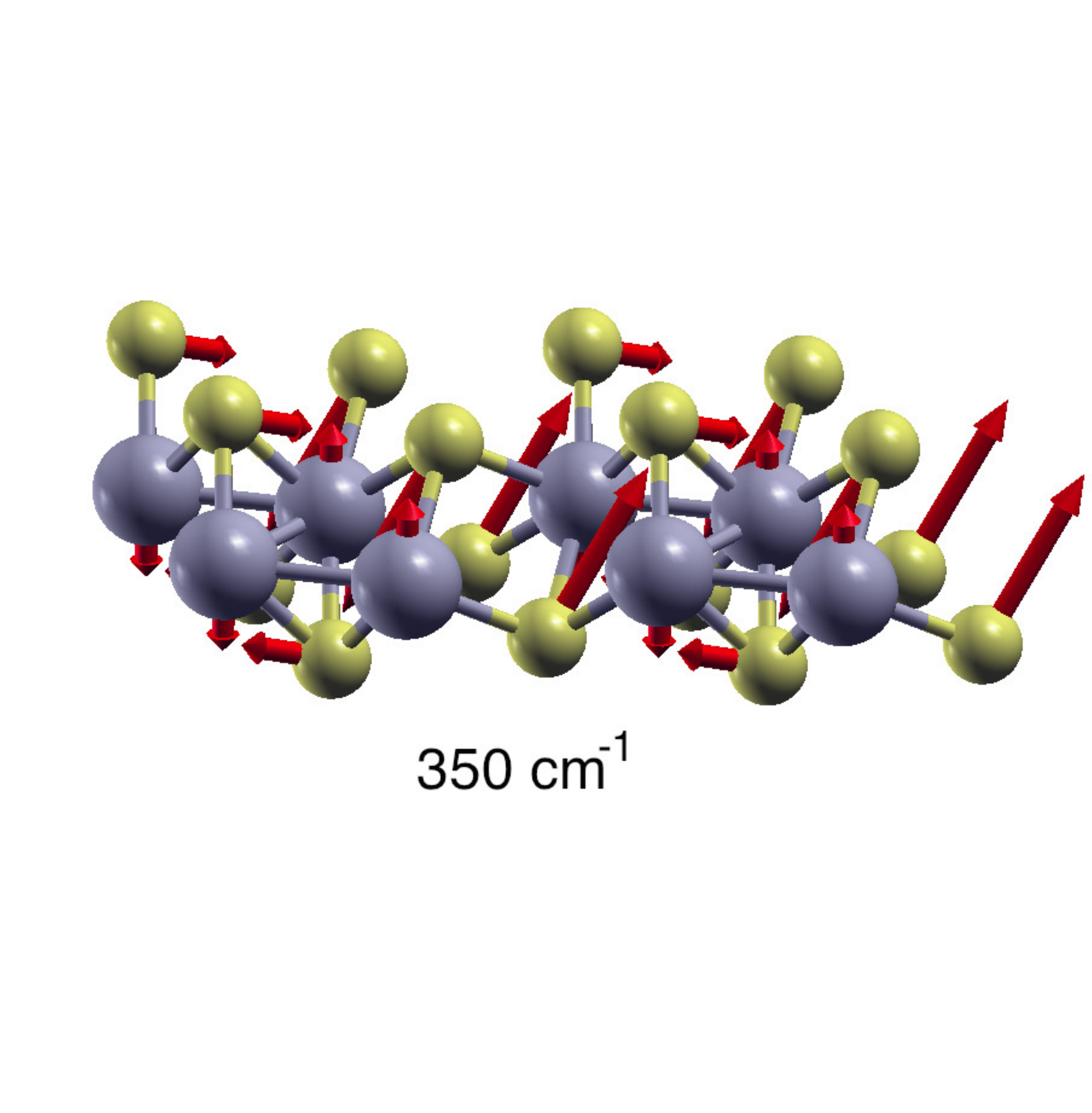}\includegraphics[scale=0.125,angle=0]{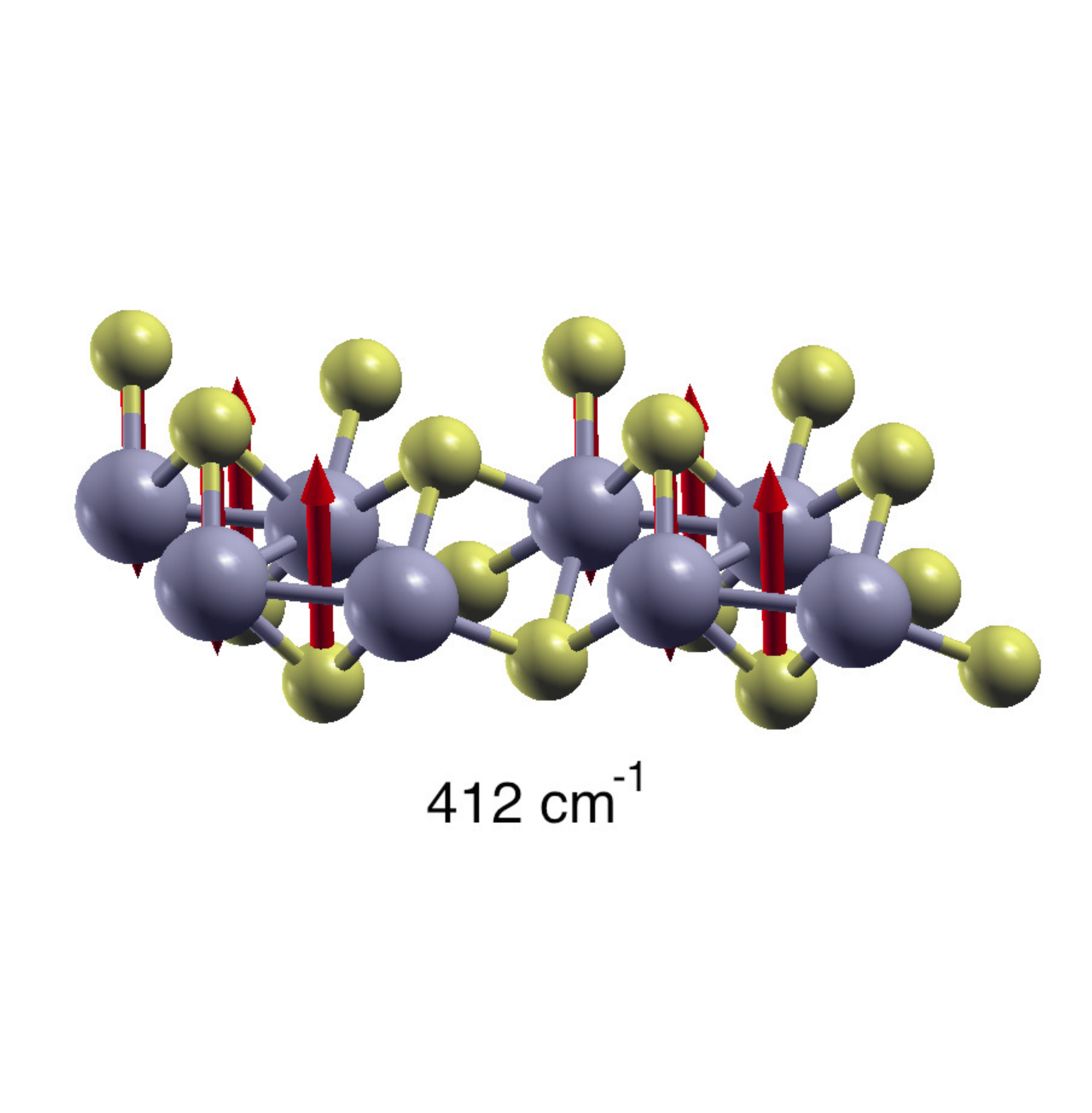}
\caption{ Raman active modes of 1T$^{\prime}$ MoS$_2$
  single-layer. The length of the arrows is proportional to the modulus of the phonon
eigenvector.}
\label{fig:Raman_Modes}
\end{figure*}

In 1H MoS$_2$, at high energy, only two Raman peaks are seen, namely the 
 E$_{\rm 2g}$ mode at $\approx 385 $ cm$^{-1}$ and the A$_{\rm 1g}$
 mode at $\approx 403$ cm$^{-1}$ (see Ref. \onlinecite{Lee}).
The experimental Raman spectra of the 1T$^{\prime}$ phase 
show two main variations with respect to H-polytypes:
(i) the E$_{\rm 2g}$ peak disappears and (ii) five additional peaks
occur (see Table \ref{tab:Raman}).
Due to the reduced symmetry of the 1T$^{'}$ structure, we do indeed find several Raman
active peaks and a very rich spectrum. The E$_{2g}$ peak is 
missing and the additional calculated Raman peaks can be associated
to the experimental ones with a high degree of accuracy.
\begin{figure}[t]
\includegraphics[scale=0.5,angle=0]{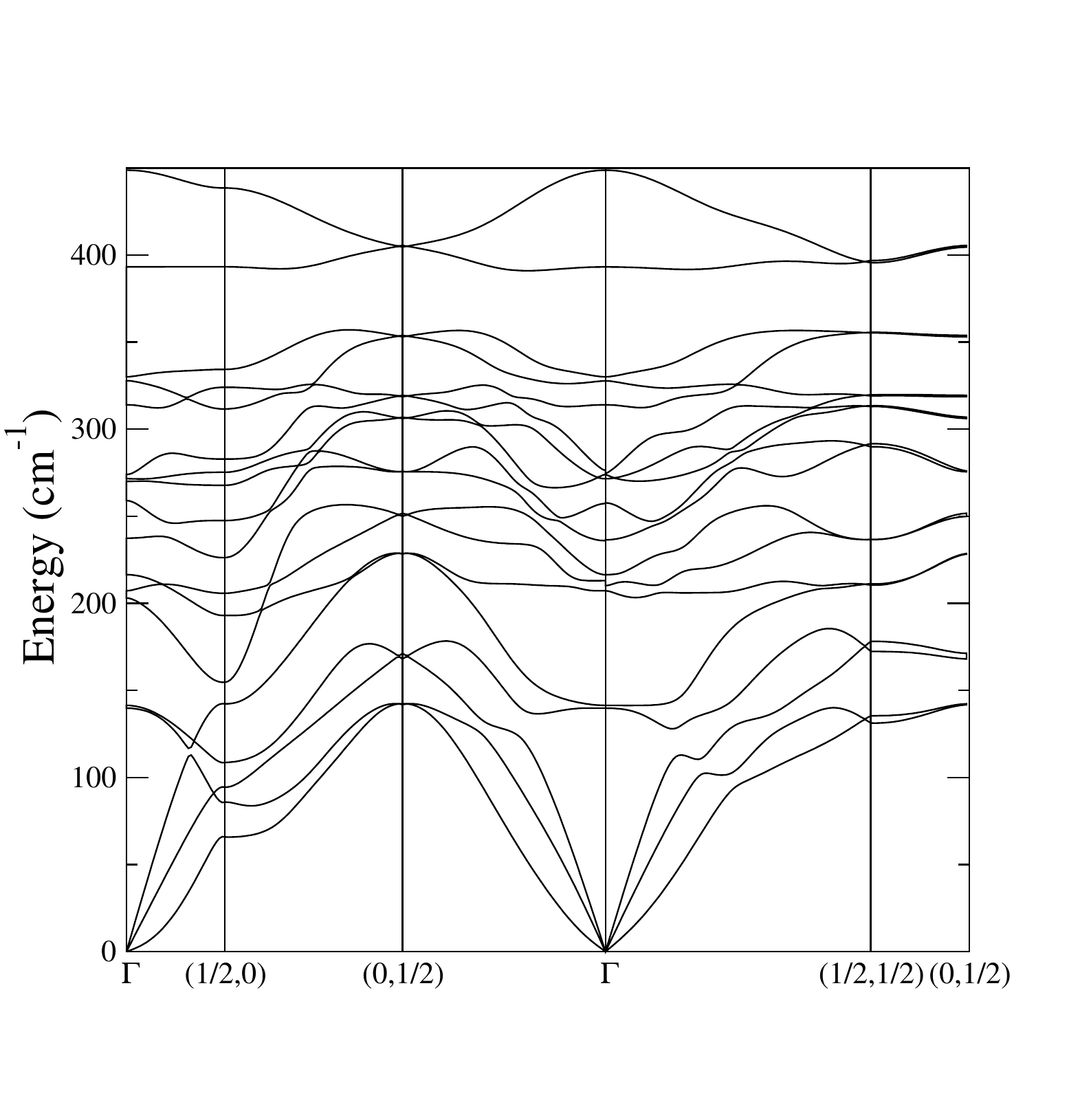}
\caption{Phonon dispersion of the 1T$^{'}$ structure along selected
directions} 
\label{fig:1Tprimeph}
\end{figure}

In our calculation the peak with the largest intensity is the so called
J$_2$ peak at $216$ cm$^{-1}$ ( $226$ cm$^{-1}$ in experiment).
This mode tends to shorten the distance
between the two zig-zag chains and to recover the 1H structure (see
Fig. \ref{fig:Raman_Modes}). In experiments \cite{Sandoval} this mode
has a much larger linewidth then all the others. This partly explains why the
experimental height of the peak is substantially reduced with respect to the Raman
intensity. 

The so called J$_1$
peak at $156$ cm$^{-1}$ in experiments is actually composed of two differents phonon
modes at $4$ cm$^{-1}$ distance one from the other. The one at $147$
cm$^{-1}$ shifts out-of-plane and in opposite directions
each stripe of Mo atoms inside the zig-zag chain.
The mode at $151$ cm$^{-1}$ is an in-plane shearing mode of one 
stripe of atom with respect to the other inside a chain. The
peaks at $233$ cm$^{-1}$ and at $286$ cm$^{-1}$
involve shifts of the S-atom layers with respect
to the Mo atoms. The J$_3$ mode at $333$ cm$^{-1}$,
in excellent agreement with experiments, tends to break
each zig-zag chain in two stripes with a slight out-of-plane component. 
The mode at $350$ cm$^{-1}$ compares favourably with the 358 cm$^{-1}$
peak detected in experiments, although in theory it has a too small
intensity. 
Finally , the mode at $412$ cm$^{-1}$ is nothing
else that the usual A$_{1g}$ mode seen in the 1H polytype. 
The agreement between the calculated zone-center energies
and the position of Raman peaks suggests that the devised structure 
closely resemble the experimental one. Some disagreement still
exists between the calculated relative intensities and the
experimental ones. However, it should be noted that Raman spectra
on different samples\cite{Sandoval,EdaNanoL} show substantially different Raman
intensities, probably due either to the inhomogeneity of
the sample composed of several phases or to the presence of adsorbates
and vacancies. 

Finally in Fig. \ref{fig:1Tprimeph} we show the calculated phonon dispersion of the 1T$^{'}$ structure
that is dynamically stable, suggesting that an energy barrier does indeed exist
between the 1H and the 1T$^{'}$ phases and that the 1T$^{'}$ is metastable.

\subsection{Catalytic adsorption\label{sec:cata}}

In order to justify the stabilization of the 1T$^{\prime}$ crystal
structure with respect to the 1H one detected in experiments, we study 
asorbates adsorption 
on the 1H, 1T , 1T$^{'}$ and 1T$^{''}$ phases .
Single layers MoS$_2$ samples at the end of the chemical exfoliation
process are fully covered with adsorbates, due to the hydratation of
Li$_x$MoS$_2$ with water.
 We focus on the simple case of 
H adsorption. We consider $4\times4$ supercells of the 
1T and 2H phase, as well as $2\times 4$ supercells of
the 1T$^{'}$ unit cell. We start considering only one H ion at random positions
on top of the MoS$_{2}$ layer and then perform several structural optimizations.
We find that the
H ion always binds to an S-atom, similarly to what
happens in WS$_2$ \cite{ChhowallaNChem}. 
Indeed, in the absence of adsorbates, a positive ( negative ) 
charge resides on the Mo (S)-atom \cite{AtacaH2O},
as it can also be inferred from the relative electronegativity of S
and Mo.
We then add  a second H  atom and find that two H 
atoms prefer to bind to different S atoms. Thus, we consider
as starting guess of the structural minimization all the possible way 
of binding H to different S atoms that are compatible with the supercell size.
\begin{figure}
\includegraphics[scale=0.35,angle=0]{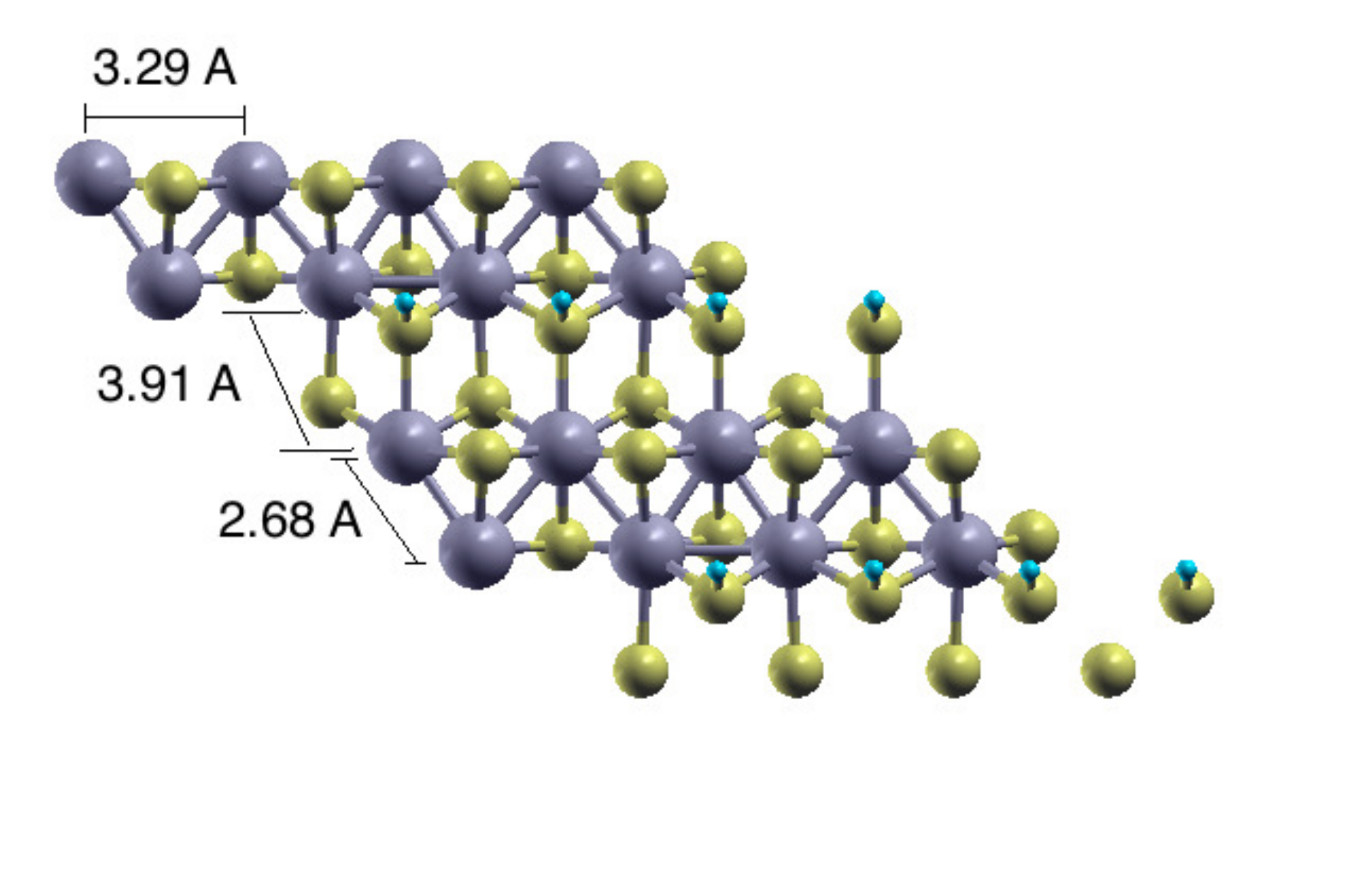}
\caption{ Most stable structure at 0.5 H coverage
   (left).  The S atoms are depicted in
yellow, while the hydrogens are the small cyan spheres.  }
\label{fig:H_Distorted}
\end{figure}

By performing structural minimization, we find that at all  H coverages
the 1H structure retains its trigonal prismatic
coordination. Similarly,
even when higher in energy, the H-covered 1T$^{\prime}$ structure never decays
into the 1H one, but preserves its zig-zag structure, although the
separation between the chains and the bonding inside the chain are affected by
H concentration. This confirms once more that an energy barrier does indeed
occur between the 1H and the 1T$^{\prime}$ structures.
Finally, we find that the  H-covered 1T structure always decays
into the  H-covered 1T$^{\prime}$ one, confirming the dynamical
instability of the 1T phase towards the 1T$^{\prime}$. At large enough
coverage,
this is also what happens to the  1T$^{''}$ structure that also decays
on the 1T$^{\prime}$.

In Fig. \ref{fig:H_Stability} we show the lowest energy configuration
of all phases  
with respect to the most stable configuration of the 1H structure at
a given H-coverage . 
We find that at H coverages superior
to $0.35$ / Mo, the 1T$^{\prime}$ phase is more stable then the 1H
one. This suggests that in chemically exfoliated MoS$_2$ monolayers, the
samples are divided in H-rich regions where the 1T$^{\prime}$
structure is stabilized and in  H-poor regions where the 1H
phase is stabilized. 

By comparing in details the 1T$^{\prime}$ structures at 0 and 0.5 H/Mo
coverage (see Fig. \ref{fig:H_Distorted}), 
it is seen that upon H adsorption the separation between the
chains strongly increases, as the shortest distance between Mo atoms
on different chains is $3.91$ \AA ( $3.71$ \AA  ) at coverage $0$ H/Mo
(0.5 H/Mo). Furthermore at coverage $0.5$ H/Mo the Mo atoms do not lay
on the same plane, as in the undistorted case, but are displaced above or below 
by $\approx 0.07$ \AA.
The increased distance between
the  chains implies a
larger band gap and more insulating character, as shown in Fig. 
\ref{fig:Dos_Distorted}.
This agrees with experiments where it was found that the
zig-zag chain structure is indeed insulating \cite{EdaACSnano, EdaNanoL}.

\section{Conclusion}

Chemically and mechanically exfoliated MoS$_2$ single-layer samples have
substantially different properties. While mechanically exfoliated
single-layers are mono-phase ( 1H phase), the chemically exfoliated
samples show coexistence of three phases, 1H, 1T and 1T$^{'}$. 
The fact that three phases experimentally coexist could lead to the 
conclusion that the three pure structures have similar energies. However, as
we have shown in the present work, this is far from being the case,
as all octahedrally coordinated phases are much higher (more then
$0.54$ eV/Mo) in energy then the trigonal prismatic one (1H). 
Moreover, the pure (i.e. without adsorbates or
vacancies) 1T phase is dynamically unstable and undergoes a phase
transition, again with 
with a considerable energy gain ($0.29$ eV / Mo), 
towards the most stable 1T$^{'}$ structure composed of separated
zig-zag chains. This finding strongly questions the detection of the
pure 1T phase in experiments \cite{EdaACSnano,EdaNanoL,Suenaga}
and points to a key role of either adsorbates or vacancies in
stabilizing the 1T metallic structure.

Wa have calculated dynamical properties of the
lowest energy octahedral structure (1T$^{'}$) and found that it is
dynamically stable, suggesting that an energy barrier does indeed exist
between the 1H and the 1T$^{'}$, similar to what happens in WS$_2$
where nudged elastic band calculations \cite{Voiry} find a $0.92$ eV/Mo barrier between the 
1T$^{'}$ and the 1H phases. By investigating catalitic adsorption on single-layer
MoS$_2$ we demonstrate the key role of adsorbates, and, more
generally, of negative charging of the MoS$_2$ layer,  in stabilizing the
1T$^{'}$ phase. This phase becomes the most stable at concentrations of
$\approx 0.35$ H / Mo.

Finally, we provided a microscopical description of the 1T$^{'}$ Raman spectrum 
attributing the J$_1$, J$_2$ and J$_3$ features to specifical
vibrations.
These features were experimentally detected in 1986
\cite{JoensenMRB1986},
but their interpretation and understanding was unknown. 

Our work represents the first complete study of static and lattice dynamical
properties of chemically exfoliated samples. We believe
that our results will be of great interest for future studies of
chemically exfoliated two dimensional crystals.

\section{Acknowledgements}
The author acknowledges useful discussions with Mannish Chhowalla and Goki
Eda. The author acknowledges support from the Graphene Flagship and
from the French state funds managed by
the ANR within the Investissements d'Avenir programme under reference 
ANR-11-IDEX-0004-02, ANR-11-BS04-0019 and ANR-13-IS10- 0003-01. 
Computer facilities were provided by CINES, CCRT and IDRIS
(project no. x2014091202).



\begin{thebibliography}{99}
\bibitem{DiSalvo} Wilson J. A., Di Salvo F. J. and Mahajan S. ,
  Advances in Physics,  {\bf 24}, 117 (1975)


\bibitem{IwasaMoS2}
J. T. Ye, Y. J. Zhang, R. Akashi, M. S. Bahramy, R. Arita
and Y. Iwasa,  Science  {\bf 30}, 1193 (2012)


\bibitem{FrindtJAP1966}
Frindt, R. F. , J. Appl. Phys.  {\bf 37}, 1928 (1966)


\bibitem{JoensenMRB1986}
Joensen, P.; Frindt, R. F.; Morrison, S. R. n
 Mater. Res. Bull.  {\bf 21}, 457 (1986)


\bibitem{FrindtPRL1972}
Frindt, R. F. , Phys. Rev. Lett., {\bf 28}, 299 (1972)


\bibitem{YangPRB1991}
Yang, D.; Sandoval, S. J.; Divigalpitiya, W. M. R.; Irwin, J. C.;
Frindt, R. F. , Phys. Rev. B, 43,
12053, (1991).


\bibitem{NovoselovPNAS}  Novoselov K. S., Jiang D., Schedin F.,
  Booth T. J., Khotkevich V. V., Morozov S. V. and  Geim A. K., 
 PNAS  {\bf 102}, 10451–10453 (2005)


\bibitem{ColemanSCI} Coleman, JN; Lotya, M; O'Neill, A et. al., 
Science,   {\bf 331} 568-571 (2011)


\bibitem{ChhowallaNChem} 
Chhowalla, M,  Shin, H. S., Eda, G.,  Li, L. J., Loh, K. P., and Zhang
H.,Nature Chemistry  {\bf 5}, 263-275 (2013)


\bibitem{Radisavljevic2011} Radisavljevic B., Radenovic A., Brivio J.,
 Giacometti V., and A. Kis,  Nature Nanotechnology {\bf 6}, 147 (2011)


\bibitem{FuhrerComment} Furer M. S. and Hone J., 
Nature Nanotechnology 2013, {\bf 8},    146–147, 
Kis A. and Radisavljevic B., 
 Nature Nanotech., {\bf 8}, 147 (2013)


\bibitem{NotAlone} {\it  Graphene is not alone}, 
    Nature Nanotechnology, {\bf  7}, 683 (2012)


\bibitem{EdaACSnano} Eda G., Fujita, T., Yamaguchi H., Voiry D., Chen M.,
and Chhowalla M., ACS Nano, {\bf 6} 7311 (2012)


\bibitem{EdaNanoL}
Eda G.,Yamaguchi H., Voiry D., Fujita T., Chen M., and Chhowalla M.,
Nano Lett.  {\bf 11}, 5111-5116 (2011)


\bibitem{Komsa} 
Komsa H. P., Kotakoski J., Kurasch S., Lehtinen O.,
Kaiser U., Krasheninnikov A. V.,
PRL {\bf 109}, 035503 (2012)


\bibitem{Suenaga}
Yung-Chang Lin, Dumitru O. Dumcenco, Ying-Sheng Huang, Kazu Suenaga,, arXiv:1310.2363


\bibitem{Lee} Lee C., Yan H., Brus L. E., Heinz T. F.,  Hone J., and
  Ryu S., ACS Nano, {\bf 4} 2695, (2010)


\bibitem{Sandoval} Sandoval S. J., Yang D., Frindt R. F., and Irwin
  J. C.,  Phys. Rev. B  {\bf 44}, 3955, (1991)


\bibitem{PBE} Perdew J. P. , Burke K., Ernzerhof M., 
Phys. Rev. Lett. {\bf 77}, 3865 (1996) 
Unused bibitems


\bibitem{QE} P. Giannozzi {\it et al.}, 
J. Phys. Condens. Matter {\bf 21}, 395502 (2009).

\bibitem{Ashcroft}
N. Ashcroft and N. D. Mermin, {\it Solid state physics}, Harcourt
College Publishers, 1976 

\bibitem{Boukhicha} Boukhicha M., Calandra M., Measson M. A., Lancry
  O., and Shukla A., 
Phys. Rev. B {\bf 87}, 195316 (2013)


\bibitem{AtacaH2O} Ataca C. and Ciraci S.,
Phys. Rev. B {\bf 85}, 195410 (2012)


\bibitem{Voiry} Voiry, Yamaguchi D. H., Li J., Silva R., Alves D. C.
  B., Fujita T., Chen M. , Asefa T., Shenoy V. B., Eda G., and
  Chhowalla M.,
Nature Materials, {\bf  12} , 850 (2013)



\end{thebibliography}
\end{document}